\documentclass{optica-article}

\journal{opticajournal} 

\articletype{Research Article}

\usepackage{lineno}
\usepackage{here} 

\usepackage{physics}
\usepackage{manyfoot}%
\usepackage{mathrsfs}%
\usepackage{xcolor}%

\usepackage{amsmath}
\usepackage{mathalpha} 
\usepackage{mathtools} 
\usepackage{hyperref} 
\usepackage{soul} 
\usepackage[table]{xcolor} 
\usepackage{colortbl} 
\usepackage[normalem]{ulem} 

\usepackage[mathcal]{eucal}
\DeclareMathAlphabet\mathbfcal{OMS}{cmsy}{b}{n}

\begin{document}

\title{Compact vacuum levitation and control platform with a single 3D-printed fiber lens}

\author{Seyed K. Alavi,\authormark{1} Jose Manuel Monterrosas Romero,\authormark{1} Pavel Ruchka, \authormark{2} Sara Jakovljević, \authormark{2} Harald Giessen, \authormark{2} and Sungkun Hong\authormark{1,*}}

\address{\authormark{1}Institute for Functional Matter and Quantum Technologies and Center for Integrated Quantum Science and Technology (IQST), University of Stuttgart, 70569 Stuttgart, Germany\\
\authormark{2}4. Physikalisches Institut, Research Center SCoPE and Center for Integrated Quantum Science and Technology (IQST), University of Stuttgart, 70569 Stuttgart, Germany}

\email{\authormark{*}sungkun.hong@fmq.uni-stuttgart.de} 

\begin{abstract*} 
Levitated dielectric particles in a vacuum have emerged as a new platform in quantum science, with applications ranging from precision acceleration and force sensing to testing quantum physics beyond the microscopic domain. Traditionally, particle levitation relies on optical tweezers formed by tightly focused laser beams, which typically require multiple bulk optical elements aligned in free space, limiting robustness and scalability of the system. To address these challenges, we employ a single optical fiber equipped with a high numerical aperture (NA) lens directly printed onto the fiber facet. This enables a compact yet robust optical levitation and detection system composed entirely of fiber-based components, eliminating the need for complex alignment. The high NA of the printed lens allows stable single-beam trapping of a dielectric nanoparticle in a vacuum, even while the fiber is in controlled motion. The high NA also allows for collecting scattered light from the particle with excellent collection efficiency, thus enabling efficient detection and feedback stabilization of the particle's motion. Our platform paves the way for practical and portable sensors based on levitated particles and provides simple yet elegant solutions to complex experiments requiring the integration of levitated particles.
\end{abstract*}

\section{Introduction}\label{sec_intro}


Optically levitated nano- and micron-sized particles in vacuum have recently emerged as a promising system for quantum science \cite{gonzalez-ballestero_levitodynamics_2021}. Levitating a particle in a high vacuum provides exceptional isolation from the environment, enabling the observation of the quantum coherent motion of the particles even at room temperature \cite{chang_cavity_2010, romero-isart_toward_2010}. It thus offers a unique opportunity to study fundamental aspects of quantum physics in previously unexplored parameter regimes \cite{romero-isart_quantum_2011}. 
This platform also holds great promise as a highly sensitive probe, with its potential applications ranging from precision sensing of force and acceleration \cite{ranjit_zeptonewton_2016, hempston_force_2017, hebestreit_sensing_2018, monteiro_force_2020, kawasaki_high_2020} to detecting dark matter and testing force laws beyond the standard model \cite{moore_searching_2021, kilian_dark_2024}.

In the past few years, the field has achieved significant milestones in controlling the motions of levitated particles at the quantum level. These include cooling of the particle’s center of mass (CoM) motion near its quantum ground state at room temperature \cite{delic_cooling_2020, magrini_real-time_2021} and under cryogenic conditions \cite{tebbenjohanns_quantum_2021}. To date, these breakthroughs have been demonstrated with standard tabletop experiments with free-space optics. Such setups require large volumes to accommodate bulky optical elements and demand high laboratory stability to maintain precise alignment. These constraints pose challenges in further scaling up the system for next-generation experiments or real-world applications, which will require higher degrees of robustness and flexibility simultaneously. As an initial step toward addressing these limitations, a mobile tweezer platform employing an optical fiber as a robust light-guiding element has been demonstrated\cite{quidant.fib.moving.apl}. However, the use of two bulky aspheric lenses aligned after the fiber limits the system’s further miniaturization, stability, and flexibility. More recently, miniaturization of tweezing optics has been achieved through microfabricated meta-lenses on flat surfaces \cite{shen_-chip_2021} and a standing wave trap based on a pre-aligned fiber assembly \cite{melo_vacuum_2024}. In particular, the latter integrates planar electrodes and the fiber assembly onto a single chip, enabling both optical trapping and electrical control within a compact, unified platform.

Here, we present the simplest form of an optical levitation platform based on a single optical fiber. Our fiber-based levitation platform consists of a single optical fiber with a high NA diffractive flat lens 3D-printed directly on the facet of the fiber\cite{two.photon.fab.fiber.ieee, two.photon.fab.nat.photonics}. This enables us to completely remove bulky optics in the system, thus reducing the size and the weight of the platform to the limits ultimately set by the bare fiber. A tight laser focus formed by the high NA printed lens allows us to trap a $142~nm$ silica nanosphere using only a single fiber without any additional counter-propagating beams. It thus entirely eliminates the need for any optical alignment. The high NA of the printed lens also enables the detection of the particle’s motion with high information collection efficiency. Using this, we achieve efficient readout and subsequent feedback cooling of the particle’s motion along the optical axis in vacuum. Similar fiber-based systems with printed high NA lenses have previously demonstrated optical trapping of microparticles\cite{Leite2018,Plidschun2021}, but not in vacuum and without readout of the motions through the same fibers. 
Finally, we demonstrate the robustness and flexibility of our system by demonstrating stable trapping of the particle while freely maneuvering the fiber.

\begin{figure}[h]
\centering
\includegraphics[]{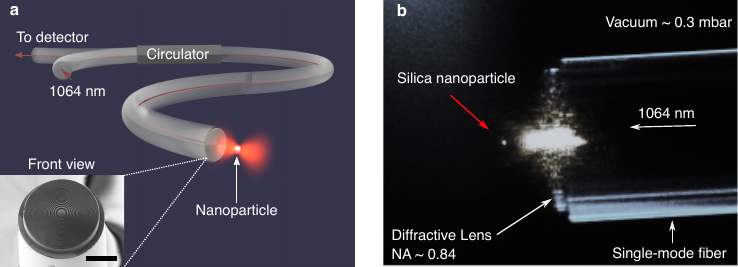}
\caption{
\textbf{Vacuum levitation of a nanoparticle using a high NA fiber lens.} \textbf{a}, Illustration of a compact, fiber-based, single-beam tweezer platform. A 1064 nm laser source is coupled to a single-mode fiber connected to the input port of a fiber optic circulator. The circulator's output port is spliced to a fiber with a diffractive lens directly printed onto the fiber facet. The laser beam exiting the fiber core is first expanded by a no-core part and reaches the diffractive Fresnel lens at the end. The lens tightly focuses the expanded beam to a focal spot at around $34~\mu m$ from the lens, forming an optical trap. Scattered light from the trapped particle is collected by the lens, coupled back into the fiber, and then redirected to a detector via the circulator, enabling efficient detection of the particle's motion. The inset depicts the scanning electron microscope image of the lens front, with a scale bar of $50~\mu m$.
\textbf{b}, Side view of the fiber tweezer captured by a commercial CMOS camera (see Fig. \ref{fig4}). A silica nanoparticle with a diameter of $142~nm$ is trapped at the focus at a pressure of $\sim 0.3~mbar$. 
}\label{fig1}
\end{figure}

\section{Results}\label{sec_res}
\subsection{Setup overview}\label{subsec_setup}
Figure \ref{fig1} illustrates our fiber-based tweezer setup comprising an optical fiber with a high NA printed lens, spliced to a commercially available fiber optic circulator (Precision Micro-Optics). 
The lens structure is fabricated on the cleaved end of a single-mode optical fiber using a two-photon polymerization technique\cite{Ruchka.QST>2022} (see Appendix \ref{app.sec2} for more details). It consists of a no-core beam expander ($550~\mu m$ in length) and a diffractive Fresnel lens\cite{Asadollahbaik2020} ($10~\mu m$ in thickness; $110~\mu m$ in diameter) with a design NA of 0.84. 
We couple an intense laser field from a high-power fiber laser (Azurlight Systems) to the input port of the circulator. The laser beam is guided through the fiber inside a vacuum chamber and focused by the lens, forming a tightly localized optical trap. Figure \ref{fig1}b shows a silica nanoparticle with a diameter of $142~nm$ stably levitated by the trap at a pressure of $0.3~mbar$. 

\subsection{Detection of the particle's motion}\label{subsec_detection}

\begin{figure}[h]
\centering
\includegraphics[]{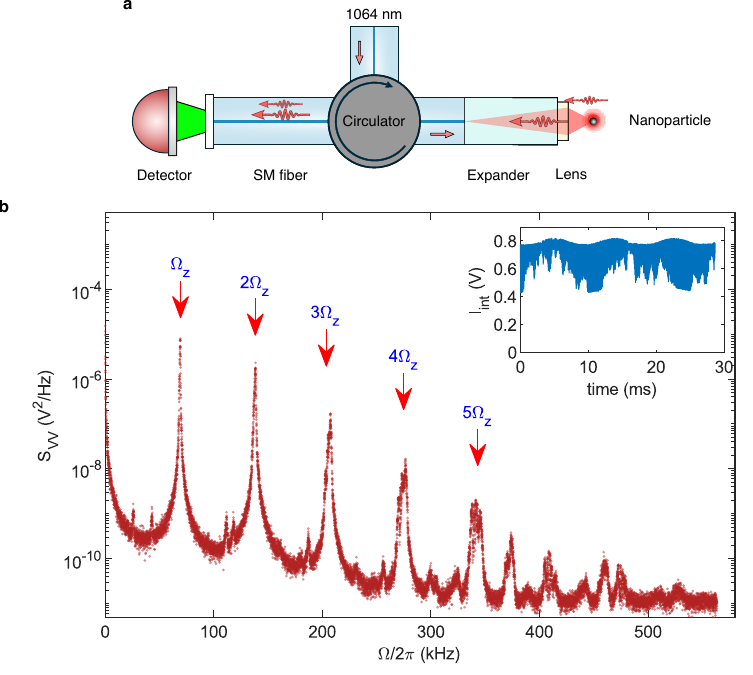}
\caption{
\textbf{Fiber-based detection of the particle's displacement.}
\textbf{a}, Schematics of the experimental platform. 
The laser field launched from the input port of the circulator and guided to the fiber lens to trap a particle. The light scattered off the particle by the trapping beam is collected by the same fiber lens and redirected to the detector (smaller wavy arrows). The particle's displacement is encoded in the phase of the scattering field. In the meantime, a small fraction of the trapping beam also reflects off the interfaces of the lens elements (larger wavy arrows) and interferes with the scattering field, serving as a local oscillator. The interfered signal is then recorded by the fiber-coupled amplified photo-detector with a gain of $1.23 \times 10^4~V/W$. 
\textbf{b}, Power spectral density (PSD) of the measured signal. The PSD shown here is obtained by averaging the PSDs of twenty individual time traces of the length $\sim 28~ ms$ (one example shown in the inset) measured consecutively. The particle's oscillatory motion along the optical axis appears as a prominent peak at $\Omega_{z}/2\pi = 69.2~kHz$ as well as its higher harmonics. 
}\label{fig2}
\end{figure}
%
%

The particle is conventionally monitored by detecting the light scattered from the particle \cite{gieseler_subkelvin_2012, quidant.fib.moving.apl, magrini_real-time_2021, tebbenjohanns_quantum_2021, shen_-chip_2021}. We follow a similar scheme to measure the particle's motion in the trap (see Fig. \ref{fig2}a). First, the lens on the fiber collects the light field back-scattered from the particle. The high NA of the fabricated lens allows for collecting the scattered field with high efficiency, thus allowing us to measure the particle's motion with high precision. The collected scattered field is redirected to a fiber-coupled amplified photo-detector via the circulator. In addition, a small fraction of the tweezer beam ($< 0.1~\%$) is reflected off the lens structure interfaces and couples back to the fiber. This reflected beam interferes with the scattering light from the particle, acting as a local oscillator with a phase reference, further simplifying our experimental apparatus. The resultant interference intensity can be modeled as, 

\begin{equation}
\label{eq_int}
\begin{aligned}
   \mathbfcal{I}_{int}(t) =  \vert \mathbfcal{E}_{scat} \vert^2 + \vert \mathbfcal{E}_{r} \vert^2 + 2\vert \mathbfcal{E}_{scat} \vert \vert \mathbfcal{E}_{r} \vert cos\bigl\{\varphi_{scat}(t) - \varphi_{r}\bigr\} + \vert \mathbfcal{E}_{r,\perp} \vert^2 \\
\end{aligned}
\end{equation}
\noindent where $ \mathbfcal{E}_{scat} = \mathbfcal{E}^0_{scat} e^{i \varphi_{scat}(t)}$ is the scattering field from the particle, $\mathbfcal{E}_{r} = \mathbfcal{E}^0_{r} e^{i\varphi_{r}}$ is the reflected field component that shares the same polarization as the scattering field, and $\mathbfcal{E}_{r,\perp}$ is the reflected field component with orthogonal polarization. 
Here, the particle's displacement $z(t)$ along the optical axis (z-axis) is directly imprinted on the phase of the scattering field $\varphi_{scat}(t)$, i.e., $\varphi_{scat}(t)=\varphi_0 + 2kz(t)$ where $k = 2\pi/\lambda$ is the wavenumber of the laser and $\varphi_0$ is the phase of the scattering field when the particle is at the trap center. Therefore, $z(t)$ can be read out by observing the intensity modulation of the interference signal. The particle's displacements perpendicular to the optical axis, on the other hand, are primarily encoded in the incident angle of the back-scattered field at the fiber interface, to which a single-mode fiber is insensitive to first order\cite{Vamivakas.OL.07,tebbenjohanns_optimal_2019}. As a result, they do not appear as strongly in our measurement. 

Figure \ref{fig2}b shows the averaged power spectral density (PSD) of the signal measured for a particle trapped at a pressure of $0.3~mbar$ by coupling a $500~mW$ laser field to the circulator input. The PSD reveals five pronounced harmonics arising from the particle’s oscillatory motion along the z-axis with the fundamental frequency of $\Omega_z/2 \pi = 69.2~kHz$. As mentioned above, the particle's oscillations perpendicular to the z-axis are not visible, as our measurement from the fiber is not sensitive to these motions. They can be separately determined by an auxiliary objective lens installed perpendicular to the fiber and are found to be $(\Omega_x/2 \pi, \Omega_y/2 \pi) = (190.3, 223)~kHz$, where $\Omega_x$ and $\Omega_y$ are the frequencies of the particle's motion along and perpendicular to the trapping beam polarization (see Appendix \ref{app.sec1}).

\begin{figure}[hbt!]
\centering
\includegraphics[]{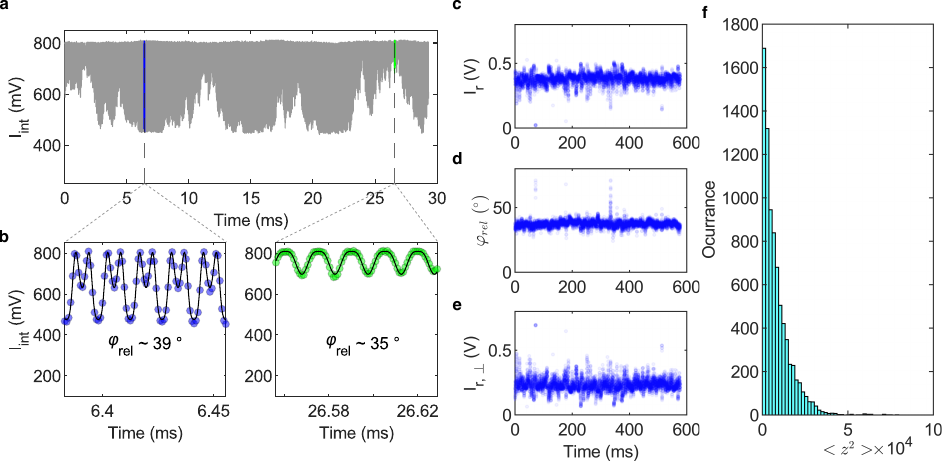}
\caption{
\textbf{Extraction of the particle dynamics from the measured signal.} 
\textbf{a}, Typical time trace of the detected signal measured at a pressure of $0.3~mbar$. The particle's displacement along the z-axis leads to modulation of the phase of the scattering field, resulting in the intensity modulation of the interference field. In addition, other parameters, such as the relative phase between the reflected field and the scattering field, also exhibit slow fluctuations, contributing to the additional fluctuation of the measured signal. \textbf{b}. Examples of the time trace segments of $\approx 72~\mu s$ in length, zoomed in from panel b at different times (shown in blue and green). Solid lines are the results of the fitting to Eq. \ref{eq_int}, assuming the particle undergoes a coherent oscillation during the period. 
\textbf{c-e}, Trends of reflected field intensity $\mathbfcal{I}_{r}$ (c), relative phase $\varphi_{rel}$ (d), and intensity of the reflected field in orthogonal polarization $\mathbfcal{I}_{r,\perp}$ (e). Fluctuations are observed for all parameters.
\textbf{f}, The distribution of the particle's displacement power $z^2$ extracted from the fittings. The distribution follows the Boltzmann probability distribution, a characteristic of the thermal state. The root-mean-square value is $\sqrt{\langle{z^2}\rangle } \approx 89.89~nm$, showing excellent agreement with the theoretical value of $\sqrt{\langle{z^2}\rangle}_{theory} \approx 89.61~nm$ obtained from the equipartition theorem.
}\label{fig3}
\end{figure}

The particle's motional frequencies obtained from the PSDs (Fig.\ref{fig2}b and Fig.\ref{figS1} in Appendix \ref{app.sec1}) and material properties can be used to estimate the characteristics of the optical trap formed by the fiber lens and the scattering response of the particle. To that end, we perform the full-field modeling of the optical tweezer \cite{Lee2025.NA} and estimate that the NA of the fiber lens, the laser power arriving at the focal spot, and the scattering power of the particle are 0.83, $135.4~mW$, and $23.49~\mu W$ respectively (see Appendix \ref{app.sec3} for more details). 

These results allow us to assess the loss of the fiber lens and the corresponding efficiency of our detection system. First, the estimated tweezer power of $135.4~mW$ is $27.1~\%$ of $500~mW$ laser input power. Considering the calibrated circulator transmission efficiency (input to fiber lens ports) of $\eta_{cir,in}=0.80$, we find that the overall efficiency of the fiber lens structure is $\eta_{fib}=0.34$. This reduced efficiency $\eta_{fib}$ is attributed to finite transmission losses of the lens structure and deviations of the diffractive Fresnel lens from an ideal lens at high NA\cite{Asadollahbaik2020}. The increased sensitivity to fabrication imperfections at high NA due to reduced feature sizes may also contribute significantly. 

The total photon collection efficiency of our detection system is then $\eta_{tot}=\eta_{NA}\cdot\eta_{fib}\cdot\eta_{cir,out}=0.064$, where $\eta_{NA}=0.24$ is the collection efficiency of the lens and $\eta_{cir,out}=0.79$ is the predetermined circulator transmission (fiber lens to output port). The resulting scattering field intensity arriving at the detector is estimated to be $\mathbfcal{I}_{scat}:=\vert \mathbfcal{E}_{scat} \vert^2=1.499~\mu W$. 

Eq. \ref{eq_int} indicates that the measurement signal also strongly depends on the static phase difference between scattered and reflected fields, $\varphi_{rel}=\varphi_0-\varphi_r$. Ideally, when $\varphi_{rel}=\pi /2$, the particle-dependent part of the signal becomes $\mathbfcal{\delta I}_{int}(t) \propto sin\bigl\{2kz(t)\bigr\} \approx 2kz(t)$ with maximal sensitivity and linearity. The strong nonlinearity, i.e., the prominent higher harmonics, shown in Figure \ref{fig2}b, however, suggests that $\varphi_{rel}$ of our measurement deviates significantly from the ideal value of $\pi /2$ (see Appendix \ref{app.sec4.sub2} for more details). 
Moreover, we observe that the measurement signal exhibits slow fluctuations in addition to fast modulations induced by the particle (see the inset in Fig. \ref{fig2}b). This is due to slow drifts in the interference parameters in Eq. \ref{eq_int}, i.e., intensities of the reflected fields ($\mathbfcal{I}_{r}:=\vert \mathbfcal{E}_{r} \vert^2$ and $\mathbfcal{I}_{r,\perp}:=\vert \mathbfcal{E}_{r,\perp} \vert^2$) as well as $\varphi_{rel}$.

To analyze the measured signal and extract the particle's motion in the presence of slowly fluctuating interference parameters, we first divide the time traces into segments with a length of $\approx 72~\mu s$. The selected time interval is short so that the parameter variations can be assumed to be negligible. Furthermore, it is also sufficiently smaller than the gas damping time expected for a given pressure ($\approx 600~\mu s$). We can, therefore, assume that the particle would undergo a coherent oscillation during this time period, i.e., $z(t)=z_{amp} \cdot cos\bigl\{\Omega_z (t-t_{0})\bigr\}$. These two assumptions allow us to fit Eq. \ref{eq_int} to the divided segments individually and determine the best-fit values for the parameters of the model. Here, the value of $\vert \mathbfcal{E}_{scat} \vert^2$ is fixed to $0.981~\mu W$ as estimated earlier. Figure \ref{fig3}b presents examples of two such segments, exhibiting excellent agreement between the measured signal and the fitted model, which confirms the validity of the method. A more detailed explanation of the fitting procedure is provided in Appendix \ref{app.sec4}.

We apply this method to a total of 7970 consecutive segments (a cumulative length of $577~ms$) to obtain the time trends for $\mathbfcal{I}_{r}$ (Fig. \ref{fig3}c), $\varphi_{rel}$ (Fig. \ref{fig3}d), and $\mathbfcal{I}_{r,\perp}$ (Fig. \ref{fig3}e), which indeed exhibit fluctuations over the total duration of time. Most importantly, we obtain the particle's motion along the z-axis, i.e., $z_{amp}$, from the same fitting results. To validate the accuracy of our method, we look at the statistical distribution of the displacement power of the particle $\langle z^2 \rangle =z_{amp}^2/2$, extracted from each segment (Fig. \ref{fig3}f). First, we confirm that it follows the Boltzmann distribution,
\begin{equation*}
P(z^2) \propto exp(- \frac{m\Omega_z^2 z^2}{2k_BT} ),    
\end{equation*}
correctly indicating that our particle is in thermal equilibrium. Moreover, the root-mean-square value of the displacement $\sqrt{\langle z^2\rangle} \approx 89.89~nm$ shows excellent agreement with the theoretical value of $\sqrt{\langle z^2\rangle_{th}}=\sqrt{k_BT/m\Omega_z^2} \approx 89.61~nm$ calculated from the equipatition theorem, where $m$ is the particle mass, $k_B$ in the Boltzmann constant, and $T = 300~K$ is the surrounding temperature. This agreement highlights the validity of our model and the reliability of our detection scheme. We further quantify the robustness of the fitting procedure using the coefficient of determination, finding 
the mean $R^2 > 0.95$ for all segments, which confirms that the extracted parameters reliably capture the particle’s motion (see \ref{app.sec4} for details).

\subsection{Feedback cooling of the particle}\label{subsec_cooling}

 \begin{figure}[hbt!]
\centering
\includegraphics[]{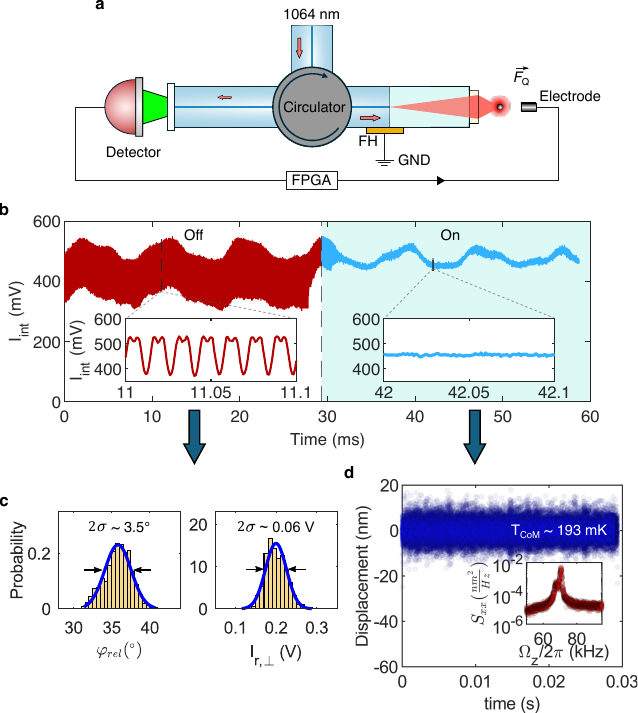}
\caption{\textbf{Cold damping of the particle's motion in high vacuum.}
\textbf{a}, Schematic representation of the fiber-based setup including the feedback scheme. The detected signal is processed with an FPGA, which is digitally filtered to generate an output feedback signal. This signal is sent to an electrode located a few millimeters from the lens. This signal creates an electric field, between the electrode and grounded fiber holder (FH) that exerts a Coulomb force proportional to the particle's position with a phase delay optimized to damp the particle’s motion along the z-axis. \textbf{b}, A time trace of the detected signal at the pressure of $1.3 \times 10^{-4} mbar$ when the feedback force is off (red) and on (blue). When the feedback is turned on, an immediate decrease in the signal amplitude is observed, indicating the corresponding decrease in the particle's oscillatory motion. The insets show exemplary zoomed-in segments with the length of $100~\mu s$ with the feedback off (left) and on (right). 
\textbf{c}, Normalized statistical distributions of $\varphi_{rel}$ (left) and $\mathbfcal{I}_{r,\perp}$ (right) extracted from the measured signal over a duration of $\sim 85~ms$ until feedback is on. The mean values of $\varphi_{rel}$ and $\mathbfcal{I}_{r,\perp}$ are used as constant parameters in Eq. \ref{eq_int} when converting the signal during the cooling phase to the particle's displacement. The standard deviations are $\sigma_{\varphi} \sim 1.73^{\circ}$ and $\sigma_{\mathbfcal{I}_{r,\perp}} \sim 30~mV$, respectively. \textbf{d}, Inferred particle's displacement during the cooling phase. The inset shows the PSD of the converted displacement signal around the particle's oscillation frequency of $\Omega_z/2\pi \approx 70.3~ kHz$. We perform the areal integration of the PSD around $\Omega_z$ to obtain the root-mean-squared value of the particle's displacement $z_{rms} = \sqrt{\langle z^2\rangle}\approx 2.26~nm$ and the corresponding effective mode temperature of $T_{CoM} = 193~mK$.  
}\label{fig4}
\end{figure}

We next demonstrate cold damping technique to cool the motion of the particles $z(t)$ by combining our fiber-based detection system with electrical force control \cite{conangla_optimal_2019, tebbenjohanns_cold_2019, magrini_real-time_2021, tebbenjohanns_quantum_2021}. Figure \ref{fig4}a illustrates the feedback control setup. The signal from the fiber-based detection system is sent to a digital controller equipped with a field programmable gate array (FPGA), where a real-time digital filter processes the signal to generate a feedback output. This signal is amplified and transmitted to an electrode located a few millimeters from the fiber lens. The electrode produces an electric field that exerts a Coulomb force on the particle proportional to the feedback signal and the particle’s charge. Within this cooling scheme, the phase delay between the measured signal and the applied force is optimized to create an effective damping term, resulting in cooling of the particle's motion (see Appendix \ref{app.sec5} for more details).

The effect of this feedback mechanism on the particle is illustrated in Figure \ref{fig4}b, which shows time traces of the detected signal at a pressure of  $1.3 \times 10^{-4}~mbar$ with feedback off (red) and on (blue). When feedback is activated, we observe the particle's motion amplitude decreases significantly, as shown in the insets. To quantitatively analyze the feedback performance, we process the measured signal in the following steps: calibration of the interference parameters from the signal before feedback activation (Step I) and conversion of the signal after feedback activation (Step II). 

In Step I, we use the method outlined in Subsection \ref{subsec_detection} to extract slowly varying interference parameters from the signal prior to feedback activation. To that end, we divide the signal trace into a total of 609 segments and fit each segment to Eq. \ref{eq_int}, yielding the trends of $\varphi_{rel}$ and $\mathbfcal{I}_{r,\perp}$ (see Appendix \ref{app.sec5.sub3} for more details). These parameters exhibit well-bounded fluctuations (see Fig. \ref{fig4}c). Here, the mean values of these parameters deviate significantly from those observed in low vacuum (Fig. \ref{fig3}). We attribute this effect to increased mechanical vibrations of the system induced by the turbo pump and the consequent vibration of the fiber-lens element, which can give rise to enhanced fluctuations in both the polarization and phase of the reflected field.

In Step II, the mean values of $\varphi_{rel}$ and $\mathbfcal{I}_{r,\perp}$ obtained from the pre-feedback phase are used as fixed parameters for processing the signal in the post-feedback phase, alongside $\mathbfcal{I}_{scat}=1.499~\mu W$. This leaves $z(t)$ and $\mathbfcal{I}_{r}$ as the remaining unknowns. Since they fluctuate on very different timescales, we can extract and process them separately by low-pass filtering the signal (see Appendix \ref{app.sec5.sub3} for more details). Fig. \ref{fig4}d presents the resulting particle displacement $z(t)$ during the cooling phase. The inset displays the PSD of the displacement signal, $S_{zz}$, in units of $ nm^2/Hz$ around the particle’s oscillation frequency ($\Omega_z/2\pi \approx 70.3~kHz$). The area under the PSD around $\Omega_z$ yields the root-mean-squared (RMS) displacement $z_{rms} = \sqrt{\langle z^2\rangle}= 2.26~nm$ and a corresponding effective mode temperature of $T_{CoM} = 193~mK$.

The PSD shown in Fig. \ref{fig4} also provides key insights into the detection and control capabilities of our system. First, the noise floor of the PSD $16.45~pm^2/Hz$ reveals a detection sensitivity of $\approx 4.05~pm/\sqrt{Hz}$, which is primarily limited by detector dark noise and classical laser noise (see Appendix \ref{app.sec5.sub5} for more details). This limitation arises because the measured signal includes a significant DC component (reflected field intensities) that inherently carries classical intensity noise. Nevertheless, the demonstrated sensitivity is approximately 10 times higher than that of the state-of-the-art fiber-based levitation platforms \cite{melo_vacuum_2024}. This is due to the significantly higher NA of the printed lens on the fiber compared to the lensless fibers used previously. 
Second, the particle’s residual displacement power at around $\Omega_z$ remains 20 dB above the detection noise floor, even with optimized feedback gain.
This observation can be attributed to the coupling of the z-motion to the uncooled x- and y-motions \cite{magrini_real-time_2021, gieseler_thermal_2013, piotrowski_simultaneous_2023} as well as effective feedback gain fluctuations induced by system parameter fluctuations in the measured signal (see Fig. \ref{figS3}). 

\subsection{Stability of the trap}\label{subsec_stability}

\begin{figure}[h]
\centering
\includegraphics[]{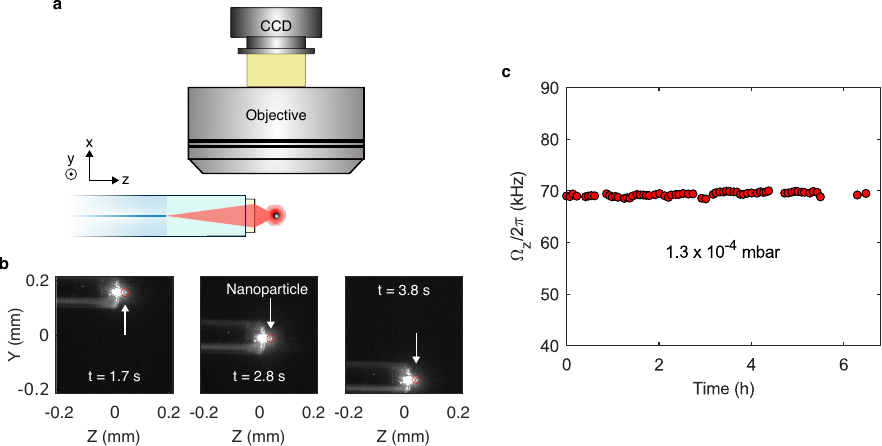}
\caption{\textbf{Stability of a fiber lens holding a levitated nanoparticle in vacuum.}
\textbf{a}, The fiber lens and the levitated particle are monitored during the transport process with an additional imaging system. Using this system, we record a video while the fiber is being moved along the y-axis.
\textbf{b}, Selected frames of a recording show the fiber and particle's position at times of $t = 1.7$, $2.8$, and $3.8~s$ (from left to right). In these frames, the particle is moved with a velocity of $\sim 130~\mu m/s$ (see Visualization 1). Arrows and red circles highlight the location of the particle in each frame. They show the particle is stably locked with the fiber while the fiber is being moved. \textbf{c}, The particle in our fiber-based trap is continuously monitored at a pressure of $1.3\times10^{-4}~mbar$ for more than six hours. During the measurement, the particle was left without any feedback-based stabilization.
}\label{fig5}
\end{figure}

Another merit of our platform is its capability to freely maneuver the lensed fiber while maintaining stable trapping of the particle. This is demonstrated by moving the fiber along, e.g., the y-axis, using a nanopositioner (Mechonics) at a pressure of $\sim 0.3~mbar$ with a speed of up to $130 ~\mu m/s$, which is the maximal speed accessible under our operating conditions (see Visualization 1). We observe that while the fiber is moved for about a distance of $0.4~mm$, the particle remains stably trapped at the focus of the fiber lens (see Fig. \ref{fig5}a). We confirm the stability of the trap by repeatedly moving the fiber at the same speed in different directions without loss of the particle (see, e.g., Visualization 2).

We further investigate the stability of our trap by monitoring the particle in a trap at a pressure of $1.3\times10^{-4}~mbar$ (see Fig. \ref{fig5}c). We observe that the particle is maintained in the trap for more than 6 hours without feedback cooling, until we intentionally lose the particle by shutting off the laser. This demonstrates excellent stability of our fiber-based optical trap based on a high NA printed lens.

These demonstrations highlight the distinct features of our compact fiber-based levitation platform based on a 3D-printed diffractive lens. First, the possibility of mounting the system on a nanopositioner and performing in situ fine positioning arises from its ultralight design, which is realized entirely with a single fiber. Moreover, the ability to translate the system without particle loss results from the use of a high NA diffractive lens to realize an alignment-free single-beam trap, combined with the capacity to sustain sufficiently high laser power to ensure trap stability even under high-vacuum conditions.

This capability expands the applicability of our system to advanced experiments that require the precise integration of various modules. For instance, our system will provide a simple yet robust solution for experiments that demand precise in-situ alignment between optical tweezers and a Paul trap \cite{conangla_extending_2020, bonvin_hybrid_2024} or micro- or nano-cavities \cite{magrini_near-field_2018, alavi_enhanced_2024}.

\section{Conclusion and Discussion}\label{sec_conclusion}
In summary, we have presented a vacuum levitation platform consisting of a single optical fiber. A high NA lens directly printed onto the fiber tip allowed us to simultaneously achieve robust levitation of a dielectric nanoparticle and efficient motion detection without the need for any alignment. The stability of the levitation was further confirmed by maintaining the trapped particle without loss while moving the fiber over several hundred micrometers, with a maximum speed available. Additionally, we verified the long-term stability of our platform in high vacuum conditions for several hours. Furthermore, we use the high detection sensitivity offered by our system to demonstrate feedback cooling of the particle's motion along the fiber's optical axis down to hundreds of millikelvin.

As previously discussed, the extent of cooling achieved was limited by several factors. First, our current fiber-based detection system is sensitive only to motion along the z-axis, restricting feedback cooling to this direction. This limitation inherently impacts cooling performance, as the cooled mode can be reheated through nonlinear coupling with uncooled motion in the x- and y-directions \cite{magrini_real-time_2021, piotrowski_simultaneous_2023, gieseler_thermal_2013}. To address this, detecting and cooling these perpendicular motions is essential. This can be achieved by introducing another fiber with a printed high NA lens oriented perpendicular to the first, which will allow for detecting the scattered fields associated with motions along the x- and y-axes.
The modular nature of our fiber-based levitation platform will ensure that key functions of the system, including alignment-free trapping and z-motion detection, remain unaffected by this addition.

The next limiting factor is our system's detection sensitivity, which is currently dominated by classical laser noise. A practical solution is to implement a balanced homodyne detection scheme with an additional local oscillator path. This approach suppresses classical laser noise via common-mode noise rejection, enabling shot-noise-limited detection.
Finally, the laser reflections from the lens elements should also be minimized, as they not only raise the shot-noise level but also hinder the precise optimization of the homodyne interferometer phase relative to the signal-carrying scattered field. Applying anti-reflection coatings\cite{Ristok.ar.coating.22} to the lens element interfaces offers a straightforward solution to this issue, realizing quantum-limited detection of the particle's motion.

The fundamental limit of feedback cooling is given by the measurement efficiency $\eta^{*}$ of a detection scheme. Specifically, the achievable minimum phonon occupation is expressed as $n_{min}\approx~(1/\sqrt{\eta^{*}}-1)/2$ \cite{wiseman_quantum_2009, wilson_measurement-based_2015}. We note that $\eta^{*}$ differs from the photon collection efficiency $\eta_{tot}$ as the information about the particle's position is imprinted differently depending on the angle of radiation \cite{tebbenjohanns_optimal_2019}. In our case, while the collection efficiency of our lens is only $\eta_{NA}=0.24$, the measurement efficiency for the particle's motion along the z-axis reaches $\eta^{*}_{NA}=0.82$. Consequently, the total measurement efficiency for the particle's z-motion amounts to $\eta^{*}_{tot}=\eta^{*}_{NA}\cdot\eta_{fib}\cdot\eta_{cir,out}\cdot\eta_{d}=0.19$, where $\eta_{d}=0.85$ is the detector's quantum efficiency. The projected $n_{min}$ is approximately 0.65, which is within a factor of two of the values achieved in state-of-the-art feedback cooling experiments \cite{magrini_real-time_2021, tebbenjohanns_quantum_2021}. The primary limiting factor is the loss from the fiber lens, with $\eta_{fib}=0.34$. We anticipate that this can be improved in the future by optimizing the lens design and fabrication process. A two-fold increase in $\eta_{fib}$ would bring $\eta^{}_{tot}$ in line with the state-of-the-art values \cite{magrini_real-time_2021, tebbenjohanns_quantum_2021}. This improvement, combined with the demonstrated robustness and flexibility of our system, will pave the way for the development of a new vacuum levitation platform that combines quantum-limited control with versatility, enabling next-generation levitodynamics experiments.

\appendix
\renewcommand{\thesection}{Appendix \Alph{section}}
\section{Detection of the particle motion perpendicular to the fiber axis}
\label{app.sec1}
\begin{figure}[H]
\centering
\includegraphics[]{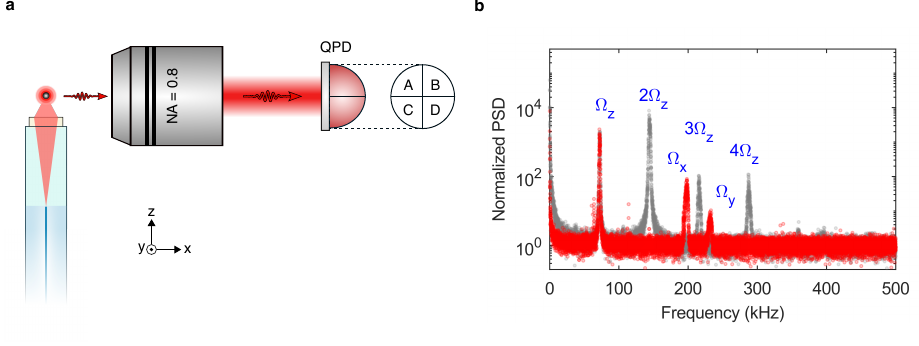}
\caption{\textbf{a}, Setup schematics for detecting full 3D motion of the particle. In addition to the fiber-based detection scheme presented in the main text, the motion of the particle is also monitored by an auxiliary high NA objective lens (NA = 0.8) installed vertically relative to the fiber. By adjusting the fiber position with a 3D nanopositioner, the foci of the fiber lens and the objective lens are aligned. The light scattered off the particle is collected by the objective and directed to a quadrant photodiode detector (QPD).
\textbf{b}, Power spectral densities (PSDs) of the signals measured by the fiber (gray) and the objective (red). Unlike the PSD obtained from the fiber lens, the PSD from the QPD clearly reveals frequency peaks corresponding to the motions perpendicular to the z-axis ($\Omega_x/2\pi=190.3~kHz$ and $\Omega_y/2\pi=223~kHz$).
}\label{figS1}
\end{figure}

\section{Fabrication of the printed lens on the fiber}
\label{app.sec2}
The high NA diffractive Fresnel lens is fabricated by two-photon polymerization 3D-printing\cite{two.photon.fab.nat.photonics} directly on a fiber tip. Prior to the printing, a no-core fiber (FG125LA, Thorlabs GmbH) is spliced to single-mode fiber (1060XP, Thorlabs GmbH) and cleaved to a length of $550~\mu m$ using Vytran GPX3800 automated glass processor (Vytran, UK). After this, we activate the fiber tip surface using Oxygen-plasma pen (Relyon PiezoBrush PZ3) to increase the adhesion of printed structure to the fiber.

For the fabrication of the lens we use Nanoscribe Photonic Professional GT 3D-printer, with the Small Features Set (IP-Dip photopolymer, 63x printing objective; Nanoscribe GmbH, Germany). The lens is fabricated with the slicing and hatching of $0.1~\mu m$, laser power of 22 \% and scan speed of $15000~\mu m/s$. After printing, we developed the lens using mr-Dev600 (micro resist technology GmbH) for 15 min and rinsed in 2-isopropanol for 3 min.

We note that this process results in a high fabrication yield and reproducibility of such 3D-printed structures. As we have recently reported \cite{bauer_side-looking_2025,imiolczyk_ultracompact_2024,zhao_lateral_2024}, these microoptics can achieve Strehl ratios exceeding 0.9 (90 \% diffraction-limited) with deviations from the design below 10 \%, and essentially 100 \% yield.

\begin{figure}
\centering
\includegraphics[]{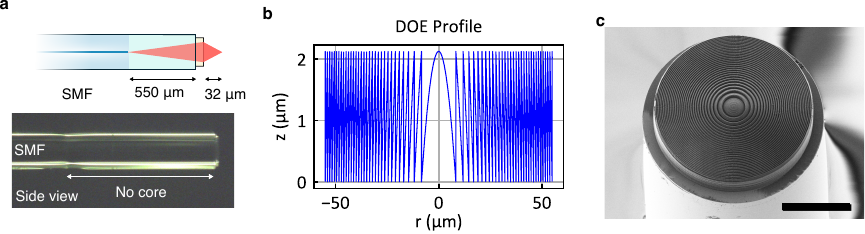}
\caption{\textbf{a}, Side-view schematic and optical image of the printed lens structure. A $550~\mu m$-long no-core glass fiber is spliced to a single-mode (SM) fiber, allowing the beam exiting the SM fiber to expand. The lens is printed at the other end of the no-core glass and focuses the beam at a working distance of $32~\mu m$.
\textbf{b}, Radial profile of the diffractive optical element (DOE), i.e., the diffractive Fresnel lens.
\textbf{c}, Scanning electron microscopy (SEM) image of the printed lens. The scale bar is $50~\mu m$.
}\label{figS2}
\end{figure}


\section{Extracting the tweezer parameters from the particle frequencies}
\label{app.sec3}
A three-dimensional profile of an optical potential is distinctively determined by how tightly the tweezer field is focused, i.e., the NA of the lens, as well as the polarization. This results in unique ratios for the trap stiffnesses, or trap frequencies, along different spatial directions.  Therefore, the ratios of the measured particle frequencies, $\Omega_x/\Omega_z$ and $\Omega_y/\Omega_z$, can be used to infer the NA of the lens and the polarization of the light field. Once the NA and the polarization are determined, the absolute value of the frequency, along with the material properties of the silica particle (e.g., refractive index $n=1.45$ and density $\rho= 1850~kg/m^3$, specified by the vendor, microparticles GmbH), can determine the intensity of the light field at the focus. Recently, the full-field modeling of the light field has been thoroughly studied in this context and successfully demonstrated its ability to accurately predict the optomechanical behavior of the particle in the optical tweezer\cite{Lee2025.NA}. We use the available code package (\href{https://github.com/moosunglee/NAeffect}{https://github.com/moosunglee/NAeffect}) and find that when the NA of the lens is 0.83 and the polarization is (0.93, $i$0.37), the theoretically expected frequency ratios are ($\Omega_x/\Omega_z$, $\Omega_y/\Omega_z$) = (2.74, 3.20), which agree well with the experimental value of (2.75, 3.22). With this result, we estimate the power of the light field arriving at the focus to be $P = 135.4~mW$. Finally, the scattering power of the particle can be calculated by the following equation\cite{Lee2025.NA}:
\begin{equation}
\label{eq_s1}
\begin{aligned}
P_{scat} = \frac{3V^2k^4P}{\pi^2\Omega_x\Omega_y} \left( \frac{n^2-1}{n^2+1} \right) ^2,
\end{aligned}
\end{equation}
where $V=4\pi r^3/3$ is the volume of the particle and $k=2\pi/\lambda$ is the wavenumber of the tweezer beam. By plugging in the numbers obtained above, as well as $r=71~nm$ and $\lambda=1064~nm$, we estimate $P_{scatt}=23.49~\mu W$.

\section{Interpretation of measured signal and particle readout}
\label{app.sec4}
\subsection{Model of measured signal based on interference effects}
\label{app.sec4.sub1}
As already discussed in the main text, the signal measured through the fiber lens can be explained by the interference between the field scattered off the particle and the field reflected by the fiber lens structure. Specifically, the field intensity measured by the detector can be described by the following equation:
\begin{equation}
\label{eq_s2}
\begin{aligned}
   \mathbfcal{I}_{int}(t) =  \mathbfcal{I}_{r} + \mathbfcal{I}_{scat} + 2\sqrt{\mathbfcal{I}_{r}\mathbfcal{I}_{scat}} cos\bigl\{\varphi_{scat}(t) - \varphi_{r}\bigr\} + \mathbfcal{I}_{r,\perp}, \\
\end{aligned}
\end{equation}
where $\mathbfcal{I}_{scat}=\vert \mathbfcal{E}_{scat} \vert^2$ and $\varphi_{scat}(t)$ are the intensity and the phase of the particle scattering field, $\mathbfcal{I}_{r}=\vert \mathbfcal{E}_{r} \vert^2$ and $\varphi_{r}$ are the intensity and the phase of the reflected field component with its polarization parallel to the scattering field, and $\mathbfcal{I}_{r,\perp}=\vert \mathbfcal{E}_{r,\perp} \vert^2$ is the intensity of the reflected field component with orthogonal polarization. Here, the particle's displacement $z(t)$ along the z-axis is encoded in $\varphi_{scat}(t)$ because it alters the photon's round-trip distance by $2z(t)$. Consequently, $\varphi_{scat}(t)$ is expressed as
\begin{equation}
\label{eq_s3}
\begin{aligned}
   \varphi_{scat}(t)=\varphi_{0}+2kz(t),
\end{aligned}
\end{equation}
where $\varphi_{0}$ is the phase of the scattering field when the particle is at the trap center.
Combining Eqs. \ref{eq_s2} and \ref{eq_s3}, we obtain the expression for the measured signal as a function of the particle’s displacement:
\begin{equation}
\label{eq_s4}
\begin{aligned}
   \mathbfcal{I}_{int}(t) =  \mathbfcal{I}_{r} + \mathbfcal{I}_{scat} + 2\sqrt{\mathbfcal{I}_{r}\mathbfcal{I}_{scat}} cos\bigl\{2kz(t)+\varphi_{rel}\bigr\} + \mathbfcal{I}_{r,\perp}, \\
\end{aligned}
\end{equation}
where $\varphi_{rel}=\varphi_{0}-\varphi_{r}$ is the relative phase between the scattering field and the reflected field when the particle is positioned at the trap center.

When the signal is observed for a duration sufficiently shorter than the particle’s damping time, the particle undergoes a coherent harmonic oscillation, $z(t)=z_{amp}cos\bigl\{\Omega_z(t-t_0)\bigr\}$, where $\Omega_z$ is the particle's oscillation frequency along the z-axis. Therefore, in this limit, the monitored signal can be approximated as
\begin{equation}
\label{eq_s5}
\begin{aligned}
   \mathbfcal{I}_{int}(t) =  \mathbfcal{I}_{r} + \mathbfcal{I}_{scat} + 2\sqrt{\mathbfcal{I}_{r}\mathbfcal{I}_{scat}} cos\left[ 2kz_{amp}cos\bigl\{\Omega_z(t-t_0)\bigr\}+\varphi_{rel}\right] + \mathbfcal{I}_{r,\perp}. \\
\end{aligned}
\end{equation}

\subsection{Relative phase and the signal nonlinearity}
\label{app.sec4.sub2}
Eq. \ref{eq_s4} indicates that the signal sensitivity to the particle’s displacement highly depends on the relative phase $\varphi_{rel}$. The maximal sensitivity $\left| \partial{\mathbfcal{I}_{int}}/\partial{z} \right|_{max}=4k\sqrt{\mathbfcal{I}_{r}\mathbfcal{I}_{scat}}$ is achieved with the optimal relative phase of $\varphi_{rel}=\pi/2$. However, in our experiment, $\varphi_{rel}$ is determined by the structure of the lens system and deviates substantially from the optimal value. This leads to not only a suboptimal sensitivity but also pronounced signal nonlinearity. The effect of $\varphi_{rel}$ on the signal nonlinearity can be best understood when expanding Eq. \ref{eq_s5} into a sum of Bessel functions according to the Jacobi-Anger identity, as
\begin{equation}
\label{eq_s6}
\begin{aligned}
   \mathbfcal{I}_{int}(t) \propto sin\left( \varphi_{rel} \right) \sum_{n=1}^{\infty}(-1)^n J_{2n-1}\left( 2kz_{amp}\right)cos\bigl\{(2n-1)\Omega_z(t-t_0)\bigr\} \\+ cos\left( \varphi_{rel} \right) \sum_{n=1}^{\infty}(-1)^n J_{2n}\left( 2kz_{amp}\right)cos\bigl\{2n\Omega_z(t-t_0)\bigr\},
\end{aligned}
\end{equation}
where $J_n(x)$ is the n-th Bessel function of the first kind. This decomposition shows that $\varphi_{rel}$ directly influences the nonlinearity by suppressing or enhancing the odd and even harmonics and vice versa. For instance, when $\varphi_{rel}=0~or~\pm \pi$, the first-order harmonics $cos\bigl\{\Omega_z(t-t_0)\bigr\}$ and all the other higher-order odd harmonics terms vanish, resulting in strong signal nonlinearity with only even harmonics.

\subsection{Extracting the particle motion from the measurement}
\label{app.sec4.sub3}
As discussed in the main text, we extract the information about the particle’s motion along the z-axis by (1) splitting the measured signal trace into individual segments with a short time interval and (2) fitting the individual segments to Eq. \ref{eq_s5}. The length of the segments is chosen to be sufficiently shorter than the damping time of the particle at a given pressure so that the particle’s motion is assumed to oscillate coherently.

Eq. \ref{eq_s5} consists of a total of seven parameters: $\mathbfcal{I}_{r}$, $\mathbfcal{I}_{scatt}$, $\mathbfcal{I}_{r,\perp}$, $z_{amp}$, $\Omega_z$, $t_0$, and $\varphi_{rel}$. $\mathbfcal{I}_{scatt}$ can be estimated a priori from a full-field simulation of the particle’s scattering response and $\Omega_z$ is deduced from the power spectral density of the measured signal (Fig. \ref{fig2}b in the main text). This leaves Eq. \ref{eq_s5} with five parameters to be determined by the fitting process. By performing fitting of the equation to individual segments of the measured signal with a short enough time interval (e.g., $\approx72~\mu s$ for the fittings presented in Fig. \ref{fig3}), we extract the coarse-grained time evolution of the particle’s oscillation amplitude $z_{amp}$, as well as fluctuating field parameters like $\mathbfcal{I}_{r}$,$\mathbfcal{I}_{r,\perp}$, and $\varphi_{rel}$.

The quality of the fits is quantified using the coefficient of determination,
\begin{equation*}
R^2 = 1 - \frac{\text{SSE}}{\text{SST}},
\end{equation*}
where SSE is the sum of squared residuals and SST is the total variance of the data. we obtain a mean of $R^2 > 0.95$ ( for fit results shown in Fig.~\ref{fig3}c), confirming that the extracted parameters reliably capture the particle's motion. These results validate the robustness of the fitting procedure and support the analysis of the particle's displacement statistics presented in the main text.

\noindent \textbf{Determination of initial guesses:} Determining reasonable initial guesses for fit parameters is an important first step of the fitting process, as it significantly influences the reliability and convergence of the fit. We estimate the initial guess values of the parameters using methods specific to each parameter. For instance, the estimates for $z_{amp}$ and $\varphi_{rel}$ can be obtained by comparing the strengths of the harmonics in the frequency domain. Eq. \ref{eq_s6} suggests that the peak ratio of the 1st-order harmonics and the 3rd-order harmonics of a Fourier-transformed trace segment is $J_1(2kz_{amp})/J_3(2kz_{amp})$, which is a function that solely depends on $z_{amp}$. By numerically solving for the roots of the given function, we can obtain the estimate for $z_{amp}$. Once $z_{amp}$ is estimated, we can extract an estimate for $\varphi_{rel}$ by taking the ratio of the 1st-order and 2nd-order harmonics from the frequency-domain data and equating it with the expression expected from Eq. \ref{eq_s6}: $J_1(2kz_{amp})/J_3(2kz_{amp})\cdot tan(\varphi_{rel})$. With the obtained guesses of $z_{amp}$ and $\varphi_{rel}$, we can subsequently estimate initial guesses of $\mathbfcal{I}_{r}$ from the amplitude of the signal oscillation, $\mathbfcal{I}_{r,\perp}$ from the DC offset of the signal, and $t_0$ from the periodicity of the trance.

\noindent \textbf{The case of small motion amplitude:} The procedure described above, however, does not always yield sound results, particularly when the particle’s displacement amplitude is small. In such cases, the nonlinearity of the signal is suppressed, making the harmonics-based estimation of $z_{amp}$ and $\varphi_{rel}$ described above effectively unusable. This can also be understood when we perform a Taylor expansion of Eq. \ref{eq_s5} in the limit of weak particle motion:
\begin{equation}
\label{eq_s7}
\begin{aligned}
   \mathbfcal{I}_{int}(t) \approx \mathbfcal{I}_{int}|_{z=0} + \left. \frac{\partial \mathbfcal{I}_{int}}{\partial z}\right| _{z=0}\cdot z = \underbrace{\left[ \mathbfcal{I}_r+\mathbfcal{I}_{scat}+2\sqrt{\mathbfcal{I}_r\mathbfcal{I}_{scat}}cos\left( \varphi_{rel} \right) \right]}_{\coloneqq DC} \\
   - \underbrace{\left[ 4k\sqrt{\mathbfcal{I}_r\mathbfcal{I}_{scat}}sin\left( \varphi_{rel} \right)\cdot z_{amp}\right]}_{\coloneqq Amp}  cos\bigl\{\Omega_z(t-t_0)\bigr\}.
\end{aligned}
\end{equation}
After the expansion, we find that the equation is reduced to one with only three effective parameters:
$DC$, $Amp$ and $t_0$. Thus, the equation becomes over-defined with five parameters.

We address this problem by taking a two-step fitting procedure. First, we fit all signal segments to Eq. \ref{eq_s5} individually using the initial guesses obtained by the method described above. We then look at the individual fitting results and evaluate $res/z_{amp}^2$, where $res$ is the residual of the fit. $res/z_{amp}^2$ measures how accurate the fitting is, given it is rescaled with the amplitude of the signal. The fitting results with small $res/z_{amp}^2$ can thus be considered trustworthy. We identify the segment with the smallest $res/z_{amp}^2$ call it the ‘best-fit’ segment. In the second step, we repeat the fitting process, starting from the segments adjacent to the ‘best-fit’ segment. Here, we feed $\mathbfcal{I}_{r}$, $\mathbfcal{I}_{r,\perp}$, and $\varphi_{rel}$ of the ‘best-fit’ segment as initial guesses. This is justified by the assumption that $\mathbfcal{I}_{r}$, $\mathbfcal{I}_{r,\perp}$, and $\varphi_{rel}$ fluctuate on a timescale much slower than the length of the segment. We then advance to the next adjacent segments and perform the fitting, feeding $\mathbfcal{I}_{r}$, $\mathbfcal{I}_{r,\perp}$, and $\varphi_{rel}$ obtained from the previous segments as initial guesses. This iterative process continues until we encounter segments where the $res/z_{amp}^2$ obtained from the first step is smaller than that of the newly obtained fit. In such cases, we retain the original first-step results as trustworthy and propagate them as initial guesses for the next adjacent segments. 

\section{Feedback cooling}
\label{app.sec5}
\subsection{Principle}
\label{app.sec5.sub1}
In a simple feedback scheme based on a delay circuit\cite{wilson_measurement-based_2015,Rossi.2018.feedback.membrane}, the measurement on the position of the particle $z(t)$ is used to generate a feedback force proportional to it with a tunable delay $\tau$, i.e., $f_{fb}(t)\sim g\cdot z(t-\tau)$ with $g$ representing a gain factor. The equation of the motion of the particle in the presence of the feedback force can be written as,
\begin{equation}
\label{eq_s8}
\begin{aligned}
   \ddot{z} + \gamma_{m}\dot{z}+\Omega_m z=\frac{1}{m}\left( f_{th}(t)-f_{fb}(t) \right)=\frac{1}{m}\left( f_{th}(t)-g\cdot z(t-\tau) \right),
\end{aligned}
\end{equation}
where $\gamma_m$ is mechanical damping resulting from gas collation for the particle with the mass of $m$, $\Omega_m$ is the oscillation frequency of the motion, $f_{th}$ is the stochastic thermal force with $\langle f_{th}(t) \rangle =0$ and $\langle f_{th}(t)f_{th}(t') \rangle=\xi \delta (t-t')$. Here $\xi=2m\gamma_mk_BT$, where $T$ is the environment temperature, is determined according to the fluctuation-dissipation theorem\cite{Kubo.1966.fluc.diss.theorem}. We take the Fourier transformation of Eq. \ref{eq_s8} to obtain the following equation:
\begin{equation}
\label{eq_s9}
\begin{aligned}
-\omega^2Z(\omega)-i\omega\gamma_mZ(\omega)+\Omega_m^2Z(\omega)=\frac{1}{m}\left( F_{th}(\omega)-ge^{-i\omega\tau}Z(\omega) \right),
\end{aligned}
\end{equation}
which then can be rearranged for $Z$:

\begin{equation}
\label{eq_s10}
\begin{aligned}
Z(\omega)=\left[ \left( \Omega_m^2-\omega^2 \right)-i\omega\gamma_m+(g/m)e^{-i\omega\tau} \right]^{-1}\left( F_{th}(\omega)/m \right).
\end{aligned}
\end{equation}
Eq. \ref{eq_s10} suggests that the response of $Z(\omega)$ peaks around $\Omega_m$ and rapidly approaches zero as it moves away from it. Therefore, the term $e^{-i\omega\tau}$ in the equation can be approximated as $e^{-i\Omega_m\tau}$. When $\tau=\pi/2\Omega_m$, $e^{-i\omega\tau}=e^{-i\pi/2}=-i$. Thus, Eq. \ref{eq_s10} becomes

\begin{equation}
\label{eq_s11}
\begin{aligned}
Z(\omega)\approx\left[ \left( \Omega_m^2-\omega^2 \right)-i\omega\left(\gamma_m+(g/m\omega) \right)\right]^{-1}\frac{F_{th}(\omega)}{m} \\
\approx \left[ \left( \Omega_m^2-\omega^2 \right)-i\omega\left(\gamma_m+\gamma_{fb} \right)\right]^{-1}\frac{F_{th}(\omega)}{m},
\end{aligned}
\end{equation}
where $\gamma_{fb}=g/m\omega \approx g/m\Omega_m$, which can now be interpreted as an additional effective damping term induced by feedback. Taking the power spectral density,

\begin{equation}
\label{eq_s12}
\begin{aligned}
S_{zz}(\omega)=\frac{2k_BT}{m}\frac{\gamma_m}{\left( \Omega_m^2-\omega^2 \right) + \left( \left(\gamma_m+\gamma_{fb} \right)\omega \right)^2 }.
\end{aligned}
\end{equation}
Finally, the effective temperature of the motion is,

\begin{equation}
\label{eq_s13}
\begin{aligned}
T_{CoM}=\frac{m\Omega_m^2\langle z^2 \rangle}{k_B}=\frac{m\Omega_m^2}{k_B}\int_{-\infty}^{\infty}\frac{S_{zz}(\omega)d\omega}{2\pi}=T\frac{\gamma_m}{\gamma_m+\gamma_{fb}}.
\end{aligned}
\end{equation}
Therefore, feedback cooling reduces the temperature of the motion from the environmental temperature by a factor of $\gamma_m/(\gamma_m+\gamma_{fb})$. 

\subsection{Implementation}
\label{app.sec5.sub2}
We implement the feedback cooling scheme described above by using the digital filter implemented with a commercial digital controller (Red Pitaya) equipped with a field programmable gate array (FPGA). Specifically, we use the Python-based open-source software interface (PyRPL; \href{https://github.com/pyrpl-fpga/pyrpl}{https://github.com/pyrpl-fpga/pyrpl}) to filter the measured signal around the particle frequency $\Omega_z$. This can be understood with Eq. \ref{eq_s6}; when we filter only the first harmonics in the equation, we obtain

\begin{equation}
\label{eq_s14}
\begin{aligned}
\mathbfcal{I}_{int,filt}(t) \propto sin\left( \varphi_{rel} \right)J_1(2kz_{amp})cos\bigl\{ \Omega_z(t-t_0)\bigr\}.
\end{aligned}
\end{equation}
Here, $J_1(2kz_{amp})$ is a nonlinear function of $z_{amp}$. This makes the gain of the filtered signal effectively dependent on the amplitude of the particle’s motion. However, when the cooling takes place and the particle motion is reduced, $J_1(2kz_{amp})$ becomes approximately a linear function of $z_{amp}$, i.e., $J_1(2kz_{amp})\approx kz_{amp}$. We then apply a gain and an appropriate time delay to the filtered signal and send it as the output signal to the electrode placed near the particle, producing an electrical feedback force on the particle.

\subsection{Extracting the particle motion from the measurement: feedback cooling case}
\label{app.sec5.sub3}

\begin{figure}
\centering
\includegraphics[]{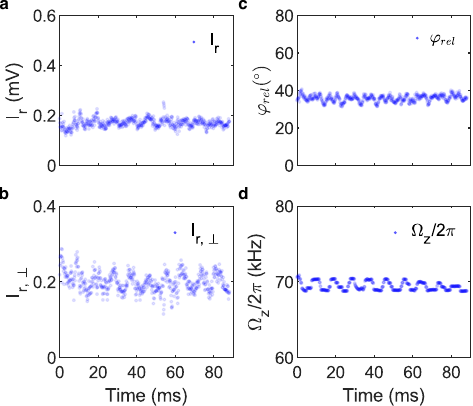}
\caption{Trends of fit parameters extracted from the measured signal before the cooling is activated at the pressure of $1.3\times 10^{-4}~mbar$. Here, $\Omega_z$ is also included as a fit parameter. All the parameters exhibit increased fluctuations with oscillatory patterns. 
We attribute this to an increased mechanical vibration of the system due to the turbo pump and its coupling to the fiber.
The mean coefficient of determination for the fits is $R^2 > 0.99$.
}\label{figS3}
\end{figure}

In the case of converting a feedback-cooled signal, the fitting method described in the earlier section cannot be directly used for several reasons. First, the cooling greatly suppresses the motion amplitude of the particle. This prevents us from unambiguously determining $z_{amp}$, along with other parameters, from the fit. In medium pressure ($~3~mbar$), as explained in Appendix \ref{app.sec4.sub1}, we solved this issue by extracting the trustworthy values of $\mathbfcal{I}_{r}$, $\mathbfcal{I}_{r,\perp}$, and $\varphi_{rel}$ from the fit results of adjacent high-amplitude signal segments and feeding the results as initial guesses. However, in the case of the cooling experiment, the cooling is activated for a duration significantly longer than the fluctuation timescale of the parameters ($\sim300~ms$). Therefore, using the parameters extracted before or after the cooling as initial guesses throughout the entire cooling period is not appropriate. Moreover, we observe that the degrees of fluctuations in $\mathbfcal{I}_{r}$, $\mathbfcal{I}_{r,\perp}$, and $\varphi_{rel}$ before cooling is activated are substantial compared to the experiments performed at higher pressures (see Fig. \ref{figS3}). To address this issue, we first extract statistical distributions of $\mathbfcal{I}_{r,\perp}$ and $\varphi_{rel}$ from the fitting results obtained during the pre-cooling phase (see Fig. \ref{fig4}c, cooling off). We assume that these parameters follow the same statistical distribution throughout the cooling phase. Next, for all possible pairs of $\mathbfcal{I}_{r,\perp}$ and $\varphi_{rel}$, we convert the signal during the cooling period to the particle’s motion using the following equation derived from Eq. \ref{eq_s7}:
\begin{equation}
\label{eq_s15}
\begin{aligned}
z(t)\approx \frac{\left( \mathbfcal{I}_r(t)+\mathbfcal{I}_{scat}+\mathbfcal{I}_{r,\perp}+2\sqrt{\mathbfcal{I}_r\mathbfcal{I}_{scat}}cos\left( \varphi_{rel}\right) \right)-I_{int}(t)}{4k\sqrt{\mathbfcal{I}_r\mathbfcal{I}_{scat}}sin\left( \varphi_{rel}\right)}.
\end{aligned}
\end{equation}
Here, all parameters on the right-hand side of the equation are known a priori ($\mathbfcal{I}_{scat}$) or assumed fixed ($\mathbfcal{I}_{r,\perp}$ and $\varphi_{rel}$) except for $\mathbfcal{I}_{r}$. We extract $\mathbfcal{I}_{r}$ separately by low-pass filtering $\mathbfcal{I}_{int}$ with a cut-off frequency of $1~kHz$. Fig. \ref{fig4}d in the main text shows the result of signal conversion and its power spectral density (PSD) when $\mathbfcal{I}_{r,\perp}$ and $\varphi_{rel}$ are fixed to the mean values of their statistical distributions. The variance of the motion $\langle z^2 \rangle$ and the corresponding mode temperature $T_{CoM}$ are obtained by integrating the PSD around the frequency of the motion and by calculating $T_{CoM}=m\Omega_z^2\langle z^2 \rangle/k_B$, respectively.

\subsection{Estimation of the cooling level}
\label{app.sec5.sub4}

\begin{figure}[h]
\centering
\includegraphics[]{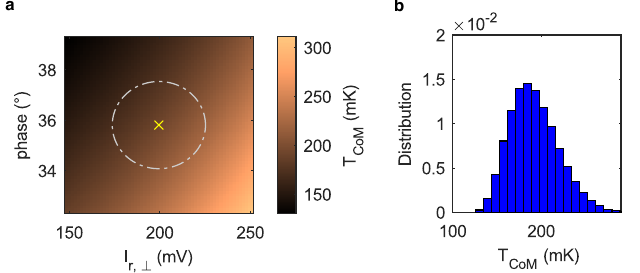}
\caption{\textbf{a}, Estimated $T_{CoM}$ as a function of $\mathbfcal{I}_{r,\perp}$ and $\varphi_{rel}$. A cross and a dashed circle represent $\left( \langle \mathbfcal{I}_{r,\perp} \rangle,\langle \varphi_{rel} \rangle \right)$ and the area within their standard deviations, respectively, of the statistical distributions obtained from the measurement before the cooling.
\textbf{b}, Estimated statistical distribution of $T_{CoM}$ calculated from the distributions of $\mathbfcal{I}_{r,\perp}$ and $\varphi_{rel}$, as well as the result from \textbf{a}.
}\label{figS4}
\end{figure}

Fig. \ref{figS4} shows the calculated $T_{CoM}$ as a function of different values of $\mathbfcal{I}_{r,\perp}$ and $\varphi_{rel}$ as well as the statistical distribution of $T_{CoM}$ estimated from the statistical samples of $\mathbfcal{I}_{r,\perp}$ and $\varphi_{rel}$ shown in Fig. \ref{fig4}c. From this, we estimate the effective temperature of the particle's motion during the cooling to be $T_{CoM}=193\pm27~mK$.

\subsection{Detection sensitivity}
\label{app.sec5.sub5}

\begin{figure}[h]
\centering
\includegraphics[]{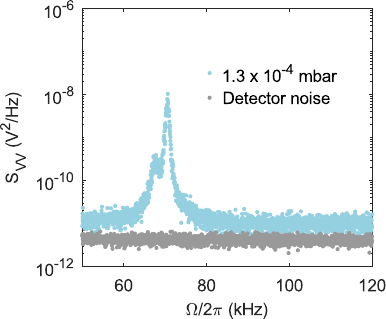}
\caption{Averaged power spectral densities of the raw signal measured during the cooling (light blue) and the detector’s dark noise (gray).
}\label{figS5}
\end{figure}

In the main text, the displacement sensitivity of $\sim4.05~pm/\sqrt{Hz}$ is estimated from $S_{zz}$ (Fig. \ref{fig4}d in the main text). Fig. \ref{figS5} shows the PSD of the raw measurement signal before conversion to the displacement, revealing the noise level of $9.43 \times 10^{-12}~V^2/Hz$. We also obtained the PSD of the detector’s bare signal and determined its dark noise level to be $4.59 \times 10^{-12}~V^2/Hz$. The difference between the noise level of the real signal and the dark noise arises from classical laser intensity noise. We also note that the shot noise for the given incoming light intensity is estimated to be $1.84 \times 10^{-15}~V^2/Hz$, three orders of magnitude lower than the measured classical noise. This indicates that the detection sensitivity of our setup could, in principle, be improved down to $\sim56~fm/\sqrt{Hz}$ with an anti-reflection coating of the fiber lens structures and a proper balanced homodyne detection setup.

\begin{backmatter}
\bmsection{Funding}
This research was supported by the Carl-Zeiss-Stiftung (Johannes-Kepler Grant through IQST, EndoPrint3D, QPhoton HoloQ), Deutsche Forschungsgemeinschaft (Projektnummers: 523178467, 431314977/GRK2642), and Bundesministerium für Bildung und Forschung (QR.X, QR.N, Integrated 3D Print).


\bmsection{Disclosures}
The authors declare no conflicts of interest.

\bmsection{Data Availability}
 The data that support this study are available upon reasonable request from the authors.

\bmsection{Supplemental material}
See Visualization 1 and 2 (videos) for supporting content. 

\end{backmatter}

\bibliography{sample}

\end{document}